\documentclass[12pt]{article}
\usepackage[dvips]{color}
\usepackage{amssymb}
\usepackage{epsfig}

\usepackage[dvips]{color}
\usepackage{epsfig}
\usepackage{amsmath}
\usepackage{graphicx}
\def\be {\begin{equation}}
\def\ee {\end{equation}}
\def\ba {\begin{eqnarray}}
\def\ea {\end{eqnarray}}

\def\k  {\kappa}

\def\L  {\Lambda}

\def\s {\sigma}

\def\bi {\begin{itemize}}
\def\ei {\end{itemize}}

\begin{document}

\title{\bf  {Interacting non-minimally coupled canonical, phantom and
quintom models of holographic dark energy in non-flat universe}}
\author{\normalsize{M. R. Setare$^{1,2}$\thanks{%
E-mail: rezakord@ipm.ir}  \, and \,Alberto Rozas-Fern\'{a}ndez
 $^{3}$\thanks{%
E-mail:a.rozas@cfmac.csic.es} }\\
\newline
\\
{\normalsize \it $^1$ Department of Science of Bijar, University of Kurdistan, Bijar, Iran}\\ { \normalsize \it $^2$ Research
Institute for Astronomy and Astrophysics of Maragha}\\
{\normalsize \it(RIAAM), Maragha,
         Iran}
\\
{\normalsize \it $^{3}$ Colina de los Chopos, Instituto de
F\'{i}sica Fundamental,}
\\
{\normalsize \it Consejo Superior de Investigaciones
Cient\'{i}ficas, Serrano 121, 28006 Madrid, Spain } }
\date{\small{}}

\maketitle
\begin{abstract}
Motivated by our recent work \cite{set1}, we generalize this work
to the interacting non-flat case. Therefore in  this paper we deal
with canonical, phantom and quintom models, with the various
fields being non-minimally coupled to gravity, within the
framework of interacting holographic dark energy. We employ the
holographic model of interacting dark energy to obtain the
equation of state for the holographic energy density in non-flat
(closed) universe enclosed by the event horizon measured from the
sphere of horizon named $L$.
 \end{abstract}

\newpage

\section{Introduction}\

Recent astrophysical data \cite{exp} strongly indicate that the
universe is accelerating at present. Therefore, it is of paramount
importance to explain why this is happening. Many theories have
been proposed recently that try to address this issue, which has
become the most important problem in cosmology. Although theories
that try to modify Einstein equations \cite{ordishov} constitute a
big part of these attempts, the mainstream explanation for this
problem, however, is known as theories of dark energy. The most
accepted idea indicates that a mysterious dominant component,
dubbed dark energy, which has negative pressure, leads to this
cosmic acceleration, though its nature and cosmological origin
still remain unknown.

The combined analysis of the cosmological observations suggests
that the universe consists of about $70\%$ dark energy, $30\%$
dust matter (cold dark matter plus baryons), and a negligible
percentage of radiation. Although little about dark energy is
known, we can still propose some candidates to describe it. The
most natural candidate to account for this acceleration is the
cosmological constant, by given the problems associated with it we
turn our attention to dynamical dark energy models. The dynamical
nature of dark energy, at least at an effective level, can be
originated from various fields, such as a canonical scalar field
(quintessence) \cite{quint}, a phantom field, that is, a scalar
field with a negative sign of the kinetic term \cite{phant}, or
the combination of quintessence and phantom in a unified model
named quintom \cite{quintom}. The advantage of this combined model
is that although in quintessence the dark energy equation of state
parameter remains always greater than $-1$ and in phantom
cosmology always
smaller than $-1$, in the quintom scenario it can cross $-1$.\\
The general consensus among theorists is that we can not entirely
understand the nature of dark energy until a complete theory of
quantum gravity is established \cite{Witten:2000zk}. The dark
energy problem may be then in essence a problem that belongs to
quantum gravity \cite{Witten:2000zk}. Thus, a complete theory of
quantum gravity should explain the properties of dark energy, such
that the energy density and the equation of state would be
determined definitely and uniquely. However, in spite of the fact
that we still lack a theory of quantum gravity, we can make some
attempts to probe the nature of dark energy by making use of the
holographic principle, which is thought to be present in a final
theory of quantum gravity. In particular, an interesting attempt
within the framework of quantum gravity is the holographic dark
energy (HDE) proposal
\cite{Cohen:1998zx,Hsu:2004ri,Li:2004rb,holoext}, which is
constructed in the light of the holographic principle and
therefore possesses some significant features of an underlying
theory of dark energy. The HDE model has been tested and
constrained by various astronomical observations
\cite{Huang:2004wt,obs3} as well as by the Anthropic Principle
\cite{Huang:2004mx}. Furthermore, the HDE model has been extended
to include the spatial curvature contribution, i.e. the HDE model
in non-flat space \cite{nonflat}.\\

On the other hand, it is usually assumed that both dark energy and
dark energy only couple gravitationally. However, given their
unknown nature and that the underlying symmetry that would set the
interaction to zero is still to be discovered, an entirely
independent behavior between the dark sectors would be very
special indeed. Moreover, since dark energy gravitates, it must be
accreted by massive compact objects such as black holes and, in a
cosmological context, the energy transfer from dark energy to dark
matter may be small but non-vanishing.

Furthermore, the coupling is not only likely but may even be
inevitable \cite{Brax:2006kg}. In addition, it might explain or at
least alleviate the coincidence problem \cite{Phenomenological
interaction, Amendola:2000uh}. It was found that an appropriate
interaction between dark energy and dark matter can influence the
perturbation dynamics and affect the lowest multipoles of the CMB
angular power spectrum \cite{Interaction in perturbation dynamics,
Interaction in the CMB multipoles}. Thus, it could be inferred
from the expansion history of the Universe, as manifested in
several experimental data. Moreover, it was suggested that the
dynamical equilibrium of collapsed structures such as clusters
would be modified due to the coupling between dark energy and dark
matter \cite{Bertolami:2007zm, Kesden}. There is no clear
consensus on the form of the coupling. Most studies on the
interaction between dark sectors rely either on the assumption of
interacting fields from the outset
\cite{Amendola:2000uh,Das:2005yj}, or from phenomenological
requirements \cite{Phenomenological interaction}. The coupling not
only affects the expansion history of the Universe but modifies
the structure formation scenario as well because we no longer have
$\rho_{m} \propto a^{-3}$. The aforesaid interaction has also been
considered from a thermodynamical perspective \cite{Wang:2007ak,
Pavon:2007gt} and has been shown that the second law of
thermodynamics imposes an energy transfer from dark energy to dark
matter.

A pressureless dark matter in interaction with holographic dark
energy is more than just another model to describe an accelerated
expansion of the universe. It provides a unifying view of
different models which are viewed as different realizations of the
interacting HDE model at the perturbative level
\cite{Zimdahl:2007ne}.

In the present paper we extend our recent work \cite{set1} to the
interacting HDE scenario in a non-flat universe.  We utilized the
horizon's radius $L$ measured from the sphere of the horizon as
the system's IR cut-off. The organization of our work is as
follows: In section 2 we construct the cosmological scenarios of
non-minimally coupled canonical, phantom and quintom fields, in
the framework of HDE. In section 3 we examine their behavior and
we discuss their cosmological implications. Finally, in section 4
we summarize our results.

\section{ Interacting non-minimally coupled fields of holographic dark energy in non-flat universe}

\subsection{Canonical field} \label{scalar}

We first consider a canonical scalar field with a non-minimal
coupling. This case was partially investigated in \cite{Ito}, and
recently extended in \cite{set1}. The action of the universe is
\begin{equation}
S=\int d^{4}x \sqrt{-g} \left[\frac{1}{2\k^{2}}
R-\frac{1}{2}\,\xi_{\phi}\phi^{2} R
-\frac{1}{2}g^{\mu\nu}\partial_{\mu}\phi\partial_{\nu}\phi
+\chi\cal{L}_{M}\right] \label{actioncan},
\end{equation}
where $\k^{2}$ is a gravitational constant. In this action we have
added a  canonical scalar field $\phi$, which in non-minimally
coupled to the curvature with coupling parameter $\xi_{\phi}$.
Lastly, the term $\cal{L}_{M}$ represents the matter content of
the universe and the term $\chi$ multiplying it accounts for the
interaction.

The presence of the non-minimal coupling leads to the effective
Newton's constant
\begin{equation}
8\pi
G_{eff}=\k^{2}\left(1-\xi_{\phi}\k^{2}\phi^{2}\right)^{-1}\label{Geff}\,.
\end{equation}

We proceed now to calculate the equation of state for the HDE
density when there is an interaction between the HDE
$\rho_{\Lambda}$ and Cold Dark Matter(CDM) with $w_{m}=0$. The
continuity equations for the HDE and CDM are
\begin{eqnarray}
\label{2eq1}&& \dot{\rho}_{\rm \Lambda}+3H(1+w_{\rm \Lambda})\rho_{\rm \Lambda} =-Q, \\
\label{2eq2}&& \dot{\rho}_{\rm m}+3H\rho_{\rm m}=Q
\end{eqnarray}
here $H=\dot{a}/a$ is the Hubble parameter. The interaction is
given by the quantity $Q=\Gamma \rho_{\Lambda}$ and describes a
decay of the HDE component into CDM with the decay rate given by
$\Gamma$. Taking the ratio of the two energy densities as
$r_{m}=\rho_{\rm m}/\rho_{\rm \Lambda}$, the above equations lead
to
\begin{equation}
\label{2eq3} \dot{r_{m}}=3Hr_{m}\Big[w_{\rm \Lambda}+
\frac{1+r_{m}}{r_{m}}\frac{\Gamma}{3H}\Big]
\end{equation}
Following Ref.\cite{Kim:2005at}, we define
\begin{eqnarray}\label{eff}
w_\Lambda ^{\rm eff}=w_\Lambda+{{\Gamma}\over {3H}}\;, \qquad w_m
^{\rm eff}=-{1\over r_{m}}{{\Gamma}\over {3H}}\;.
\end{eqnarray}
Thus, the continuity equations can be written in their standard
form as
\begin{equation}
\dot{\rho}_\Lambda + 3H(1+w_\Lambda^{\rm eff})\rho_\Lambda =
0\;,\label{definew1}
\end{equation}
\begin{equation}
\dot{\rho}_m + 3H(1+w_m^{\rm eff})\rho_m = 0\; \label{definew2}.
\end{equation}
We now consider the non-flat Friedmann-Robertson-Walker universe
with line element
 \be\label{metr}
ds^{2}=-dt^{2}+a^{2}(t)\left(\frac{dr^2}{1-kr^2}+r^2d\Omega^{2}\right).
 \ee
where $k$ denotes the curvature of space $k=0,1,-1$ for flat,
closed and open universe, respectively. It must be noticed that a
closed universe with a small positive curvature ($\Omega_k\sim
0.01$) is compatible with observations \cite{ {wmap}, {ws}}. As
usual, we use the Friedmann equation to relate the curvature of
the universe to the energy density. In the interacting case we are
dealing with, the Friedmann equations and the evolution equation
for the scalar field are
\begin{equation}
 H^{2}=\frac{\kappa^{2}\left(\rho_{m}+\rho_{\Lambda}+\frac{1}{2}\dot{\phi}^{2}+6\xi_{\phi}
H\phi\dot{\phi}\right)}{3\left(1-\xi_{\phi}\kappa^{2}\phi^{2}\right)}
\label{eqn4}
\end{equation}
\begin{equation}
\dot{\phi}\left[\ddot{\phi}+3H\dot{\phi}+6\xi_{\phi}\left(\dot{H}+2H^{2}\right)\phi\right]=-Q\label{eqn5}
\end{equation}
\begin{equation}
\dot{\rho}_{m}+\dot{\rho}_{\Lambda}+3H\left(\rho_{m}+\rho_{\Lambda}+p_{m}+p_{\Lambda}\right)=0\label{eqn6}.
\end{equation}
In these expressions,
 $p_{m}$ and $\rho_{m}$ are the
pressure and density of the matter content of the universe,
respectively. Finally, $p_\L$ and $\rho_\L$ are the corresponding
components of dark energy, which are attributed to the scalar
field.

We define as usual
\begin{equation} \label{2eq9} \Omega_{\rm
m}=\frac{\rho_{m}}{\rho_{cr}}=\frac{ \rho_{\rm
m}}{3M_p^2H^2},\hspace{1cm}\Omega_{\rm
\Lambda}=\frac{\rho_{\Lambda}}{\rho_{cr}}=\frac{ \rho_{\rm
\Lambda}}{3M^2_pH^2},\hspace{1cm}\Omega_{k}=\frac{k}{a^2H^2}
\end{equation}
where $M^2_p=(8\pi G_{eff})^{-1}$.

 Now we can rewrite the first
Friedmann equation as
\begin{equation} \label{2eq10} \Omega_{\rm m}+\Omega_{\rm
\Lambda}=1+\Omega_{k}.
\end{equation}
This allows us to express $r_{\rm m}$ and $r_{\rm k}=\rho_{\rm
k}/\rho_{\rm \Lambda}$ in terms of $\Omega_{\rm \Lambda}$ and
$\Omega_{\rm k}$ as
\begin{equation}
\label{2eq11} r_{\rm m}=\frac{1-\Omega_{\rm \Lambda}+\Omega_{\rm
k}}{\Omega_{\rm \Lambda}},~~r_{\rm k}=\frac{\Omega_{\rm
k}}{\Omega_{\rm \Lambda}}.
\end{equation}

In the non-flat universe, our choice for HDE density is
 \be \label{holoda}
  \rho_\Lambda=\frac{3}{\kappa^{2}}(1-\xi_{\phi}\kappa^{2}\phi^{2})L^{-2},
 \ee
where $L$ is defined as
\begin{equation}\label{leq}
 L=ar(t),
\end{equation}
here, $a$, is the scale factor and $r(t)$ can be obtained from the
following equation
\begin{equation}\label{rdef}
\int_0^{r(t)}\frac{dr}{\sqrt{1-kr^2}}=\int_t^\infty
\frac{dt}{a}=\frac{R_h}{a},
\end{equation}
where $R_h$ is the event horizon. Therefore while $R_h$ is the
radial size of the event horizon measured in the $r$ direction,
$L$ is the radius of the event horizon measured on the sphere of
the horizon. For the closed universe we have (the same calculation
is valid for the open universe by a transformation)
 \be \label{req}
 r(t)=\frac{1}{\sqrt{k}} sin y.
 \ee
where $y\equiv \sqrt{k}R_h/a$. As in \cite{set1} we are interested
in extracting power-law solutions of our cosmological model
(\ref{eqn4})-(\ref{eqn6}), in the case of a dark-energy dominated
universe ($\rho_{m},p_{m}\ll 1$). Thus, we are looking for
solutions of the form
\begin{eqnarray}
&&a(t)=a_{0}t^{r}\nonumber\\
&&\phi(t)=\phi_{0}t^{s_{\phi}}.
\end{eqnarray}
The insertion of the second ansatz allows us to express the HDE
density as
 \be \label{holoda2}
  \rho_\Lambda(t)=\frac{3}{\kappa^{2}}(1-\xi_{\phi}\kappa^{2}{{\phi_{0}}}^2t^{2s_{\phi}})L^{-2}.
 \ee

 Using the definitions
$\Omega_{\Lambda}=\frac{\rho_{\Lambda}}{\rho_{cr}}$ and
$\rho_{cr}=3M_{p}^{2}H^2$, we get

\begin{equation}\label{hl}
HL=\frac{1}{\sqrt{\Omega_{\Lambda}}}.
\end{equation}
Now using Eqs.(\ref{leq}, \ref{rdef}, \ref{req}, \ref{hl}), we
obtain
 \be \label{ldot}
 \dot L= HL+ a \dot{r}(t)=\frac{1}{\sqrt{\Omega_\Lambda}}-cos y.
\end{equation}
By considering  the definition  of Eq.(\ref{holoda2}) for the HDE
$\rho_{\rm \Lambda}(t)$, and using Eqs.(\ref{hl}, \ref{ldot}) one
can find
\begin{equation}\label{roeq}
\dot{\rho_{\Lambda}}=-2s_{\phi}t^{-1}\varrho_{\Lambda}\left(\frac{\xi_{\phi}\kappa^{2}{{\phi_{0}}}^2t^{2s_{\phi}}}{1-\xi_{\phi}\kappa^{2}{{\phi_{0}}}^2t^{2s_{\phi}}}\right)-2H\left(1-{\sqrt{\Omega_\Lambda}}\cos
y\right)\rho_{\Lambda}.
\end{equation}
If we substitute the above relation into Eq.(\ref{2eq1}) and use
the definition $Q=\Gamma \rho_{\Lambda}$, we arrive at
\begin{equation}\label{stateq}
w_{\rm
\Lambda}(t)=-\frac{8\xi_{\phi}(\xi_{\phi}\kappa^{2}{{\phi_{0}}}^2)}{t^{-2s_{\phi}}-\xi_{\phi}\kappa^{2}{{\phi_{0}}}^2}
-\left(\frac{1}{3}+\frac{2\sqrt{\Omega_{\rm \Lambda}}}{3}\cos
y+\frac{\Gamma}{3H}\right).
\end{equation}
Here as in Ref.\cite{WGA}, we choose the following relation for
the decay rate
\begin{equation}\label{decayeq}
\Gamma=3b^2(1+r_{m})H
\end{equation}
with the coupling constant $b^2$. Using Eq.(\ref{2eq11}), the
above decay rate yields
\begin{equation}\label{decayeq2}
\Gamma=3b^2H\frac{(1+\Omega_{k})}{\Omega_{\Lambda}}.
\end{equation}
Substituting this relation into Eq.(\ref{stateq}), one finds the
HDE equation of state parameter
\begin{equation} \label{3eq4}
w_{\rm
\Lambda}(t)=-\frac{8\xi_{\phi}(\xi_{\phi}\kappa^{2}{{\phi_{0}}}^2)}{t^{-2s_{\phi}}-\xi_{\phi}\kappa^{2}{{\phi_{0}}}^2}
-\frac{1}{3}\left[1+2\sqrt{\Omega_{\rm \Lambda}}\cos
y+\frac{3b^2(1+\Omega_{k})}{\Omega_{\rm \Lambda}}\right].
\end{equation}
According to relation $y\equiv \sqrt{k}R_h/a$, $\cos y=1$ when
$k=0$, so in this case $\Omega_{k}=0$ and therefore, in flat
universe, the HDE equation of state is given by
\begin{equation} \label{3eq401}
w_{\rm
\Lambda}(t)=-\frac{8\xi_{\phi}(\xi_{\phi}\kappa^{2}{{\phi_{0}}}^2)}{t^{-2s_{\phi}}-\xi_{\phi}\kappa^{2}{{\phi_{0}}}^2}
-\frac{1}{3}\left(1+2\sqrt{\Omega_{\rm
\Lambda}}+\frac{3b^2}{\Omega_{\rm \Lambda}}\right) .
\end{equation}
From Eqs.(\ref{eff}, \ref{decayeq2}, \ref{3eq4}), we have the
effective equation of state as

\begin{equation} \label{3eq402}
w_{\rm
\Lambda}^{eff}=-\frac{8\xi_{\phi}(\xi_{\phi}\kappa^{2}{{\phi_{0}}}^2)}{t^{-2s_{\phi}}-\xi_{\phi}\kappa^{2}{{\phi_{0}}}^2}
-\frac{1}{3}\left(1+2\sqrt{\Omega_{\rm \Lambda}}\cos y\right).
\end{equation}or expressed in terms of the redshift $z$ as
\begin{equation} \label{wcanonical}
w_{\rm
\Lambda}^{eff}(z)=-\frac{8\xi_{\phi}(\xi_{\phi}\kappa^{2}{{\phi_{0}}}^2)}{e^{-24\xi_{\phi}ln(1+z)}-\xi_{\phi}\kappa^{2}{{\phi_{0}}}^2}
-\frac{1}{3}\left(1+2\sqrt{\Omega_{\rm \Lambda}}\cos y\right).
\end{equation}
 On the
other hand, the effective equation of state for CDM is given
differently by
\begin{equation} \label{wmeff}
\omega^{\rm eff}_{\rm
m}(z)=-\frac{b^{2}(1+\Omega_{k})}{(1+\Omega_{k}-\Omega_{\Lambda})}.
\end{equation}
Now we are in a position to derive two coupled equations whose
solutions determine
 the effective equations of state (as in Ref.[24]).
Eq.(\ref{2eq3}) leads to one differential equation for
$\Omega_{\rm \Lambda}$
\begin{equation} \label{diffeqnlambda}
-(1+z)\frac{d \Omega_{\rm \Lambda}}{dz}=-3\Omega_{\rm
\Lambda}(1-\Omega_{\rm \Lambda}+\Omega_{\rm k})\Big(\omega^{\rm
eff}_{\rm \Lambda}-\omega^{\rm eff}_{\rm m}\Big)+\Omega_{\rm
k}\Omega_{\rm \Lambda}(1+3\omega^{\rm eff}_{\rm \Lambda}\Big).
\end{equation}
The remaining differential equation for $\Omega_{\rm k}$ comes
from the derivative of $r_{\rm k}$ in Eq.(\ref{2eq11}) using
Eq.(\ref{2eq10}) as
\begin{equation} \label{omegak}
-(1+z)\frac{d \Omega_{\rm k}}{dz}=-3\Omega_{\rm k}(1-\Omega_{\rm
\Lambda}+\Omega_{\rm k})\Big(\omega^{\rm eff}_{\rm
\Lambda}-\omega^{\rm eff}_{\rm m}\Big)+\Omega_{\rm
k}\Big(1+\Omega_{\rm k}\Big)\Big(1+3\omega^{\rm eff}_{\rm
\Lambda}\Big).
\end{equation}

In order to obtain a solution, we have to solve the above coupled
equations numerically by considering the initial condition at the
present time: $\frac{d \Omega_{\rm
\Lambda}}{dz}|_{z=0}>0,~\Omega^{0}_{\rm \Lambda}=0.73,
\Omega^{0}_{\rm k=1}=0.01$ and $\Omega^{0}_{\rm k=0}=0.0$.

\subsection{Phantom field} \label{phantom}
In this subsection we consider a phantom field with a non-minimal
coupling, that is, a field with an opposite sign in the kinetic
term in the Lagrangian \cite{phant}. Such models are widely used
in order to have $w_\L$ less than $-1$. The action of the universe
in this case is
\begin{equation}
S=\int d^{4}x \sqrt{-g} \left[\frac{1}{2\k^{2}}
R-\frac{1}{2}\xi_{\sigma}\sigma^{2} R
+\frac{1}{2}g^{\mu\nu}\partial_{\mu}\sigma\partial_{\nu}\sigma
+\chi\cal{L}_{M}\right] \label{actionphan}.
\end{equation}
In this action we have added a phantom field $\sigma$, which in
non-minimally coupled to the curvature with coupling parameter
$\xi_{\sigma}$. Lastly, the term $\cal{L}_{M}$ represents the
matter content of the universe and the term $\chi$ multiplying it
accounts for the interaction.
The presence of the non-minimal
coupling leads to the effective Newton's constant:
\begin{equation}
8\pi
G_{eff}=\k^{2}\left(1-\xi_{\sigma}\k^{2}\s^{2}\right)^{-1}\label{Geff2}\,.
\end{equation}

We shall follow a procedure similar to the one used in the
previous subsection in order to obtain the equation of state.

The cosmological equations and the evolution equation for the
interacting phantom field are given by

\begin{equation}
 H^{2}=\frac{\kappa^{2}\left(\rho_{m}+\rho_{\Lambda}-\frac{1}{2}\dot{\sigma}^{2}+6\xi_{\phi}
H\sigma\dot{\sigma}\right)}{3\left(1-\xi_{\phi}\kappa^{2}\sigma^{2}\right)}
\label{eqn4b}
\end{equation}
\begin{equation}
\dot{\sigma}\left[\ddot{\sigma}+3H\dot{\sigma}-6\xi_{\sigma}\left(\dot{H}+2H^{2}\right)\sigma\right]=-Q\label{eqn5b}
\end{equation}
\begin{equation}
\dot{\rho}_{m}+\dot{\rho}_{\Lambda}+3H\left(\rho_{m}+\rho_{\Lambda}+p_{m}+p_{\Lambda}\right)=0\label{eqn6b}
\end{equation}
and the non-flat HDE can be expressed as

\be \label{holodaphan}
  \rho_\Lambda=\frac{3}{\kappa^{2}}(1-\xi_{\sigma}\kappa^{2}\sigma^{2})L^{-2}
 \ee
where we have used the effective nature of the Newton's constant
(\ref{Geff2}).

We examine power-law solutions of equations
(\ref{eqn4b})-(\ref{eqn6b}), in the case of a dark-energy
dominated universe ($\rho_{m},p_{m}\ll 1$). Thus, we impose:
\begin{eqnarray}
&&a(t)=a_{0} t^{r}\nonumber\\
&&\sigma(t)=\sigma_{0}t^{s_{\sigma}} \label{powerlaw2}.
\end{eqnarray}
Using (\ref{holodaphan}) we have that
\begin{equation}\label{holoda3}
\rho_{\Lambda}(t)=\frac{3}{\kappa^{2}}(1-\xi_{\sigma}\kappa^{2}\sigma_{0}^{2}t^{2s_{\sigma}})L^{-2}
\end{equation}
and as in the previous subsection $L$ is defined as in
Eq.(\ref{leq}). By taking the definition  of Eq.(\ref{holoda3})
for the HDE density $\rho_{\rm \Lambda}(t)$, and making use of
Eqs.(\ref{hl}, \ref{ldot}) one can obtain
\begin{equation}\label{roeq2}
\dot{\rho_{\Lambda}}=-2s_{\sigma}t^{-1}\varrho_{\Lambda}\left(\frac{\xi_{\sigma}\kappa^{2}{{\sigma_{0}}}^2t^{2s_{\sigma}}}{1-\xi_{\sigma}\kappa^{2}{{\sigma_{0}}}^2t^{2s_{\sigma}}}\right)-2H\left(1-{\sqrt{\Omega_\Lambda}}\cos
y\right)\rho_{\Lambda}.
\end{equation}
Substitution of the above relation into Eq.(\ref{2eq1}) and use of
the definition $Q=\Gamma \rho_{\Lambda}$, yields
\begin{equation}\label{stateq2}
w_{\rm
\Lambda}(t)=\frac{8\xi_{\sigma}(\xi_{\sigma}\kappa^{2}{{\sigma_{0}}}^2)}{t^{-2s_{\sigma}}-\xi_{\sigma}\kappa^{2}{{\sigma_{0}}}^2}
-\left(\frac{1}{3}+\frac{2\sqrt{\Omega_{\rm \Lambda}}}{3}\cos
y+\frac{\Gamma}{3H}\right).
\end{equation}
Inserting Eq.(\ref{decayeq2}) into Eq.(\ref{stateq2}) allows us to
obtain the equation of state parameter
\begin{equation}\label{stateq3}
w_{\rm
\Lambda}(t)=\frac{8\xi_{\sigma}(\xi_{\sigma}\kappa^{2}{{\sigma_{0}}}^2)}{t^{-2s_{\sigma}}-\xi_{\sigma}\kappa^{2}{{\sigma_{0}}}^2}
-\frac{1}{3}\left[1+2\sqrt{\Omega_{\rm \Lambda}}\cos
y+\frac{3b^2(1+\Omega_{k})}{\Omega_{\rm \Lambda}}\right].
\end{equation}
Considering the relation $y\equiv \sqrt{k}R_h/a$, $\cos y=1$ when
$k=0$, i.e. $\Omega_{k}=0$ and therefore, in flat universe, the
HDE equation of state is expressed as
\begin{equation} \label{3eq4011}
w_{\rm
\Lambda}(t)=\frac{8\xi_{\sigma}(\xi_{\sigma}\kappa^{2}{{\sigma_{0}}}^2)}{t^{-2s_{\sigma}}-\xi_{\sigma}\kappa^{2}{{\sigma_{0}}}^2}
-\frac{1}{3}\left(1+2\sqrt{\Omega_{\rm
\Lambda}}+\frac{3b^2}{\Omega_{\rm \Lambda}}\right) .
\end{equation}
From Eqs.(\ref{eff}, \ref{decayeq2}, \ref{stateq3}), we arrive at
the effective equation of state
\begin{equation} \label{wlambda}
w_{\rm
\Lambda}^{eff}=\frac{8\xi_{\sigma}(\xi_{\sigma}\kappa^{2}{{\sigma_{0}}}^2)}{t^{-2s_{\sigma}}-\xi_{\sigma}\kappa^{2}{{\sigma_{0}}}^2}
-\frac{1}{3}\left(1+2\sqrt{\Omega_{\rm \Lambda}}\cos y\right).
\end{equation}Similarly to what was done in the previous subsection, we can express
(\ref{wlambda}) in terms of the redshift $z$ obtaining
\begin{equation} \label{wphantom}
w_{\rm
\Lambda}^{eff}(z)=\frac{8\xi_{\sigma}(\xi_{\sigma}\kappa^{2}{{\sigma_{0}}}^2)}{e^{24\xi_{\sigma}ln(1+z)}-\xi_{\sigma}\kappa^{2}{{\sigma_{0}}}^2}
-\frac{1}{3}\left(1+2\sqrt{\Omega_{\rm \Lambda}}\cos y\right).
\end{equation}
Finally, we insert Eq.(\ref{wphantom}) into
Eqs.(\ref{diffeqnlambda}) and (\ref{omegak}) and solve them
numerically by considering the initial condition at the present
time: $\frac{d \Omega_{\rm \Lambda}}{dz}|_{z=0}>0,~\Omega^{0}_{\rm
\Lambda}=0.73, \Omega^{0}_{\rm k=1}=0.01$ and $\Omega^{0}_{\rm
k=0}=0.0$.

\subsection{Quintom model} \label{quintom}
In this subsection we consider the quintom cosmological scenario
\cite{quintom}, where we consider simultaneously a canonical and a
phantom field, both with non-minimal coupling. As we have stated
in the introduction, this combined cosmological paradigm has been
shown to be capable to describe the crossing of the phantom divide
$w_\L=-1$. The action of this model is given by
\begin{eqnarray}
S=\int d^{4}x \sqrt{-g} \left[\frac{1}{2\kappa^{2}}
R-\frac{1}{2}\xi_{\phi}\phi^{2} R-\frac{1}{2}\xi_{\sigma}\sigma^{2} R-\right.\nonumber\\
\left.
-\frac{1}{2}g^{\mu\nu}\partial_{\mu}\phi\partial_{\nu}\phi+\frac{1}{2}g^{\mu\nu}\partial_{\mu}\sigma\partial_{\nu}\sigma
+\chi\cal{L}_{M}\right], \label{actionquint}
\end{eqnarray}
and the presence of the non-minimal coupling leads to the
effective Newton's constant
\begin{equation}
8\pi
G_{eff}=\kappa^{2}\left[1-\kappa^{2}(\xi_{\phi}\phi^{2}+\xi_{\sigma}\sigma^{2})\right]^{-1}\label{Geff3}\,.
\end{equation}
The cosmological equations and the evolution equation for the
canonical and phantom fields in the interacting case are the
following

\begin{equation}
H^{2}=\frac{\kappa^{2}\left(\rho_{m}+\rho_{\Lambda}+\frac{1}{2}\dot{\phi}^{2}-\frac{1}{2}\dot{\sigma}^{2}+6\xi_{\phi}H\phi\dot{\phi}+6\xi_{\sigma}H\sigma\dot{\sigma}\right)}{3\left[1-\kappa^{2}\left(\xi_{\phi}\phi^{2}+\xi_{\sigma}\sigma^{2}\right)\right]}\label{eqn4c}
\end{equation}

\begin{equation}
\dot{\phi}\left[\ddot{\phi}+3H\dot{\phi}+6\xi_{\phi}\left(\dot{H}+2H^{2}\right)\phi\right]=-Q\label{eqn5c}
\end{equation}

\begin{equation}
\dot{\sigma}\left[\ddot{\sigma}+3H\dot{\sigma}-6\xi_{\sigma}\left(\dot{H}+2H^{2}\right)\sigma\right]=-Q\label{eqn5c2}
\end{equation}

\begin{equation}
\dot{\rho}_{m}+\dot{\rho}_{\Lambda}+3H\left(\rho_{m}+\rho_{\Lambda}+p_{m}+p_{\Lambda}\right)=0\label{eqn6c}
\end{equation} as usual the non-flat HDE is given by
\begin{equation}
\rho_{\Lambda}=\frac{3}{\kappa^{2}}\left[1-\kappa^{2}\left(\xi_{\phi}\phi^{2}+\xi_{\sigma}\sigma^{2}\right)\right]L^{-2},\label{rhoL3}
\end{equation}
after making use of the effective nature of the Newton's constant
(\ref{Geff3}). We examine power-law solutions of equations
(\ref{eqn4c})-(\ref{eqn6c}), in the case of a dark-energy
dominated universe ($\rho_{m},p_{m}\ll 1$). Thus, we impose:
\begin{eqnarray}
&&a(t)=a_{0} t^{r}\nonumber\\
&&\phi(t)=\phi_{0}t^{s_{\phi}}\nonumber\\
&&\sigma(t)=\sigma_{0}t^{s_{\sigma}} \label{powerlaw3}.
\end{eqnarray}

Substituting into (\ref{rhoL3}) yields
\begin{equation}
\rho_{\Lambda}(t)=\frac{3}{\kappa^{2}}\left[1-\kappa^{2}\left(\xi_{\phi}\phi_{0}^{2}t^{2s_{\phi}}+\xi_{\sigma}\sigma_{0}^{2}t^{2s_{\sigma}}\right)\right]L^{-2}
\label{holoda4}
\end{equation}
and as in the previous subsections $L$ is defined as in
Eq.(\ref{leq}).

Using the definition  of Eq.(\ref{holoda4}) for the HDE density
$\rho_{\rm \Lambda}(t)$, and considering Eqs.(\ref{hl},
\ref{ldot}) gives us
\begin{equation}\label{roeq3}
\dot{\rho_{\Lambda}}=-2\left(s_{\phi}+s_{\sigma}\right)t^{-1}\left[\kappa^{2}\left(\xi_{\phi}\phi_{0}^{2}t^{2s_{\phi}}+\xi_{\sigma}\sigma_{0}^{2}t^{2s\sigma}\right)\right]-2H\left(1-\sqrt{\Omega_{\Lambda}}cosy\right)\rho_{\Lambda}.
\end{equation}
The substitution of the above relation into Eq.(\ref{2eq1}) and
the use of the definition $Q=\Gamma \rho_{\Lambda}$, yields
\begin{equation}\label{stateq4}
w_{\rm
\Lambda}(t)=16\left[\frac{\kappa^{2}\left(\xi_{\phi}^{2}\phi_{0}^{2}t^{2s_{\phi}}-\xi_{\sigma}^{2}\sigma_{0}^{2}t^{2s_{\sigma}}\right)}{\kappa^{2}\left(\xi_{\phi}\phi_{0}^{2}t^{2s_{\phi}}+\xi_{\sigma}\sigma_{0}^{2}t^{2s_{\sigma}}\right)-1}\right]
-\left(\frac{1}{3}+\frac{2\sqrt{\Omega_{\rm \Lambda}}}{3}\cos
y+\frac{\Gamma}{3H}\right).
\end{equation}
Inserting Eq.(\ref{decayeq2}) into Eq.(\ref{stateq4}) gives the
equation of state parameter
\begin{equation}\label{stateq5}
w_{\rm
\Lambda}(t)=16\left[\frac{\kappa^{2}\left(\xi_{\phi}^{2}\phi_{0}^{2}t^{2s_{\phi}}-\xi_{\sigma}^{2}\sigma_{0}^{2}t^{2s_{\sigma}}\right)}{\kappa^{2}\left(\xi_{\phi}\phi_{0}^{2}t^{2s_{\phi}}+\xi_{\sigma}\sigma_{0}^{2}t^{2s_{\sigma}}\right)-1}\right]
-\frac{1}{3}\left[1+2\sqrt{\Omega_{\rm \Lambda}}\cos
y+\frac{3b^2(1+\Omega_{k})}{\Omega_{\rm \Lambda}}\right].
\end{equation}
Given the relation $y\equiv \sqrt{k}R_h/a$, $\cos y=1$ when $k=0$,
i.e. $\Omega_{k}=0$ and therefore, in flat universe, the HDE
equation of state is given by
\begin{equation} \label{3eq403}
w_{\rm
\Lambda}(t)=16\left[\frac{\kappa^{2}\left(\xi_{\phi}^{2}\phi_{0}^{2}t^{2s_{\phi}}-\xi_{\sigma}^{2}\sigma_{0}^{2}t^{2s_{\sigma}}\right)}{\kappa^{2}\left(\xi_{\phi}\phi_{0}^{2}t^{2s_{\phi}}+\xi_{\sigma}\sigma_{0}^{2}t^{2s_{\sigma}}\right)-1}\right]
-\frac{1}{3}\left(1+2\sqrt{\Omega_{\rm
\Lambda}}+\frac{3b^2}{\Omega_{\rm \Lambda}}\right) .
\end{equation}
From Eqs.(\ref{eff}, \ref{decayeq2}, \ref{stateq5}), we have the
effective equation of state as
\begin{equation} \label{3eq4022}
w_{\rm
\Lambda}^{eff}=16\left[\frac{\kappa^{2}\left(\xi_{\phi}^{2}\phi_{0}^{2}t^{2s_{\phi}}-\xi_{\sigma}^{2}\sigma_{0}^{2}t^{2s_{\sigma}}\right)}{\kappa^{2}\left(\xi_{\phi}\phi_{0}^{2}t^{2s_{\phi}}+\xi_{\sigma}\sigma_{0}^{2}t^{2s_{\sigma}}\right)-1}\right]
-\frac{1}{3}\left(1+2\sqrt{\Omega_{\rm \Lambda}}\cos y\right).
\end{equation}
that can be expressed in terms of the redshift $z$ as
\begin{equation} \label{wquintom}
w_{\rm
\Lambda}^{eff}(z)=16\left[\frac{\kappa^{2}\left(\xi_{\phi}^{2}\phi_{0}^{2}e^{24\xi_{\phi}ln(1+z)}-\xi_{\sigma}^{2}\sigma_{0}^{2}e^{-24\xi_{\sigma}ln(1+z)}\right)}{\kappa^{2}\left(\xi_{\phi}\phi_{0}^{2}e^{24\xi_{\phi}ln(1+z)}+\xi_{\sigma}\sigma_{0}^{2}e^{-24\xi_{\sigma}ln(1+z)}\right)-1}\right]
-\frac{1}{3}\left(1+2\sqrt{\Omega_{\rm \Lambda}}\cos y\right).
\end{equation}
Finally, as in previous subsections, we insert Eq.(\ref{wquintom})
into Eqs.(\ref{diffeqnlambda}) and (\ref{omegak}) and solve them
numerically by considering the initial condition at the present
time: $\frac{d \Omega_{\rm \Lambda}}{dz}|_{z=0}>0,~\Omega^{0}_{\rm
\Lambda}=0.73, \Omega^{0}_{\rm k=1}=0.01$ and $\Omega^{0}_{\rm
k=0}=0.0$.
\section{Cosmological implications} \label{cosmimpl}

In the previous subsections we have obtained the equation of state
parameter of dark energy, $w_{\rm \Lambda}^{eff}(z)$, in terms of
the coupling parameters $\xi_{\phi}$, $\xi_{\sigma}$ and the
amplitudes $\phi_0$, $\sigma_0$. In the present section we
investigate the cosmological implications for each case.
\subsection{Canonical field} \label{cosmimplcan}

     In the case of an interacting canonical field, non-minimally
coupled to gravity, $w_{\rm \Lambda}^{eff}(z)$ is given by the
relation (\ref{wcanonical}). In Fig.1 we depict $w_{\rm
\Lambda}^{eff}(z)$ for two different values of the coupling
$\xi_{\phi}$ and for three different values of the combination
$\k^2\phi_0^2$. Note that the physical requirement of an expanding
universe results in an upper limit for $\xi_{\phi}$, namely
$\xi_{\phi}<1/6$ (see \cite{set1}).

As we observe, the value of $w_{\rm \Lambda}^{eff}(z)$ at $z=0$,
that is, its current value $w_{\rm \Lambda0}^{eff}$, decreases as
$\xi_{\phi}$ increases, while its dependence on $\k^2\phi_0^2$ is
non-monotonic. However, in this interacting canonical field case
$w_{\rm \Lambda0}^{eff}$ is always greater than $-1$,
independently of the values of $\xi_{\phi}$ and $\k^2\phi_0^2$.
This was expected since this case is known to be insufficient to
describe the crossing of the phantom divide $w_{\rm
\Lambda}^{eff}=-1$ from above \cite{Kim:2005at}.

Secondly, we can see that for $\k^2\phi_0^2$ of the order of 1 or
below, we obtain a divergence of $w_{\rm \Lambda}^{eff}$. This
behavior is a clear prediction of relation (\ref{wcanonical}),
since it possesses a singularity at
\begin{equation}
z_{s}=-1+\left(\xi_{\phi}\k^2\phi_0^2\right)^{-\frac{1}{24\xi_{\phi}}}
\label{singcan}.
\end{equation} Therefore, the combinations of $\xi_{\phi}$
and $\k^2\phi_0^2$ that satisfy the equation giving a positive
$z_s$ must be excluded.

 Finally, we mention that the effect of
non-flat universe is negligible as the curves for $k=+1,0$ appear
superimposed.

\subsection{Phantom field} \label{cosmimplphan}
In the case of an interacting phantom field, non-minimally coupled
to gravity, $w_{\rm \Lambda}^{eff}(z)$ is given by relation
(\ref{wphantom}). In Fig.2 we depict $w_{\rm \Lambda}^{eff}(z)$
for two different values of the coupling $\xi_{\sigma}$ and for
three different values of the combination $\k^2\s_0^2$. Note that
in this case the physical requirement of an expanding universe,
results in an upper limit for $\xi_{\sigma}$, namely
$-1/6<\xi_{\sigma}$ (see \cite{set1}).

As we can see, the value of $w_{\rm \Lambda0}^{eff}(z)$ is now a
non-monotonic function of $\xi_{\sigma}$ and $\k^2\s_0^2$.
Furthermore, we observe that for some particular combinations of
$\xi_{\sigma}$ and $\k^2\s_0^2$, as a consequence of the
singularity of (\ref{wphantom}), there is a divergence of $w_{\rm
\Lambda}^{eff}(z)$ at
\begin{equation}
z_{s}=-1+\left(\xi_{\sigma}\k^2\s_0^2\right)^{-\frac{1}{24\xi_{\sigma}}}
\label{singphan}.
\end{equation}
Thus, the combinations of $\xi_{\sigma}$ and $\k^2\s_0^2$ that
satisfy this transcendental equation giving a positive $z_s$ must
be excluded.

In the case at hand we can see that $w_{\rm \Lambda0}^{eff}(z)$ is
always greater than $-1$, independently of the values of
$\xi_{\sigma}$ and $\k^2\s_0^2$. This is expected as we cannot
have $w_{\rm \Lambda0}^{eff}(z)<-1$ for a phantom field in
interacting HDE (see \cite{Kim:2006kk}). Furthermore, the effect
of non-flat universe is negligible as the curves for $k=+1,0$
appear superimposed. Therefore, the non-flat universe cannot
induce the phantom phase even if one includes a non-minimal
coupling in the interacting HDE framework.

\subsection{Quintom model} \label{cosmimplquint}
In the case of the combined quintom model, that is, when both the
canonical and phantom fields are considered to be non-minimally
coupled to gravity simultaneously, $w_{\rm \Lambda}^{eff}(z)$ is
given by relation (\ref{wquintom}). In Fig.3 we depict $w_{\rm
\Lambda}^{eff}(z)$ for two different values of the coupling
$\xi_{\phi}$ and for three different combinations of
$\k^2\phi_0^2$ and $\k^2\s_0^2$. Note that in this case the
physical requirement of an expanding universe, results in an upper
limit for $\xi_{\phi}$, namely $\xi_{\phi}<1/6$ (see \cite{set1}).
The value of $w_{\rm \Lambda0}^{eff}(z)$ is a monotonic function
of $\xi_{\phi}$. As in the previous cases, for some particular
combinations of $\xi_{\phi}$, $\k^2\phi_0^2$ and $\k^2\s_0^2$, as
a consequence of (\ref{wquintom}), there is a singularity of
$w_{\rm \Lambda}^{eff}(z)$ at a specific $z_{s}$. The form of the
denominator of (\ref{wquintom}) does not allow for an explicit
expression of $z_{s}$, but a numerical investigation provides the
specific excluded parameter values.

As we can observe in Fig. 3, $w_{\rm \Lambda0}^{eff}(z)$ is
greater than $-1$, independently of the values of $\xi_{\phi}$,
$\k^2\phi_0^2$ and $\k^2\s_0^2$. Once again, the effect of
non-flat universe is negligible as the curves for $k=+1,0$ appear
superimposed. It turns out that not even the quintom scenario with
non-minimal interacting HDE can describe the phantom regime.

\begin{figure}[canonical]
\begin{center}
\includegraphics[width=.8\textwidth]{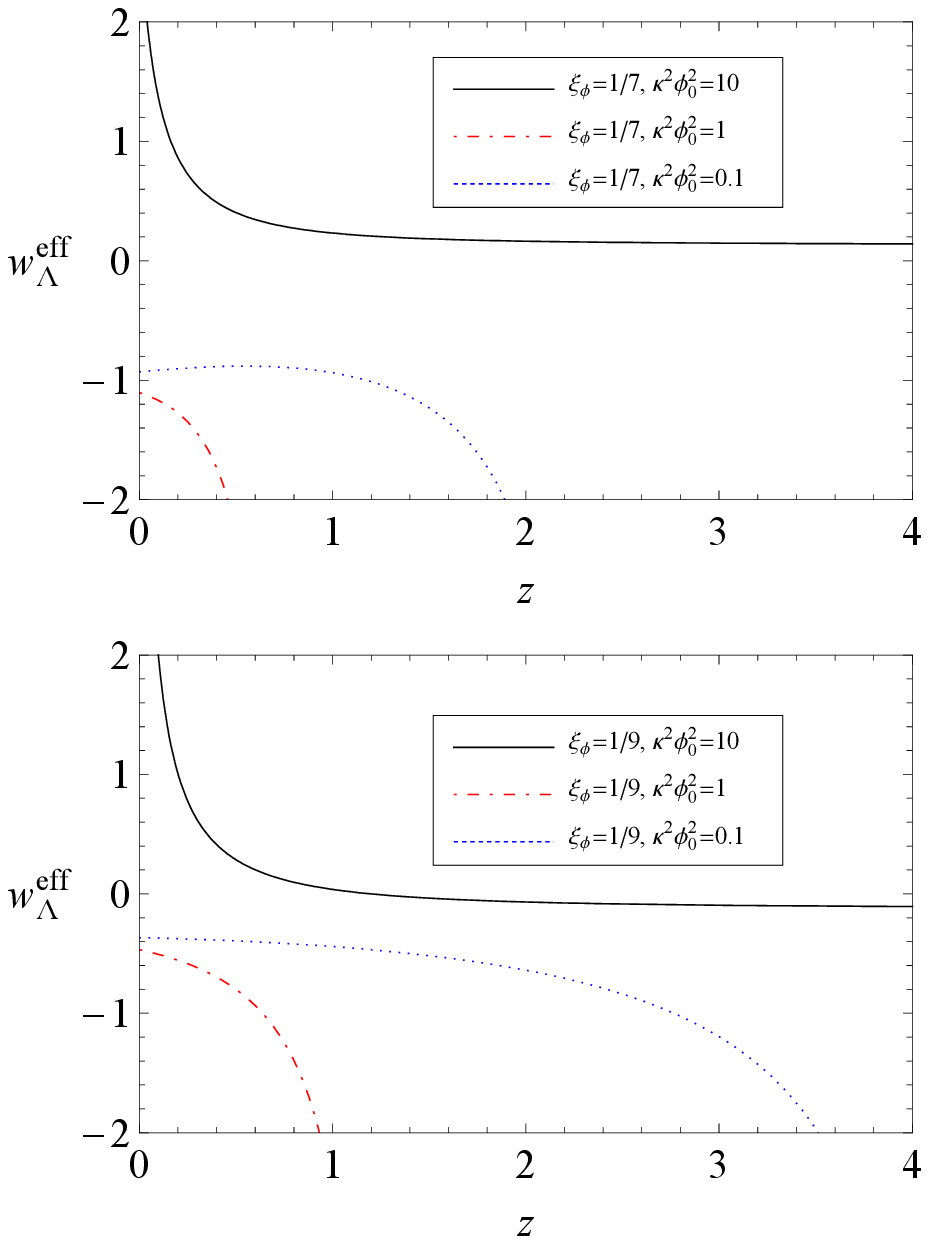}
\end{center}

\caption{{\it  $w_{\rm \Lambda}^{eff}$ vs $z$ in the interacting
canonical field case, for $\xi_{\phi}=1/7$, $\xi_{\phi}=1/9$,
$b^{2}=0.01$ and $k=+1,0$, where in each case the combination
$\k^2\phi_0^2$ is taken equal to $10$, $1$, $0.1$ respectively.
The curves for $k=+1,0$ appear superimposed showing that the
effect of the non-flat universe is negligible. The divergence of
$w_{\rm \Lambda}^{eff}$ is a direct consequence of the singularity
of (\ref{wcanonical}), and thus the corresponding combinations of
and $\k^2\phi_0^2$ must be excluded.}} \label{canonicalfig}
\end{figure}
\begin{figure}[phantom]
\begin{center}
\includegraphics[width=.8\textwidth]{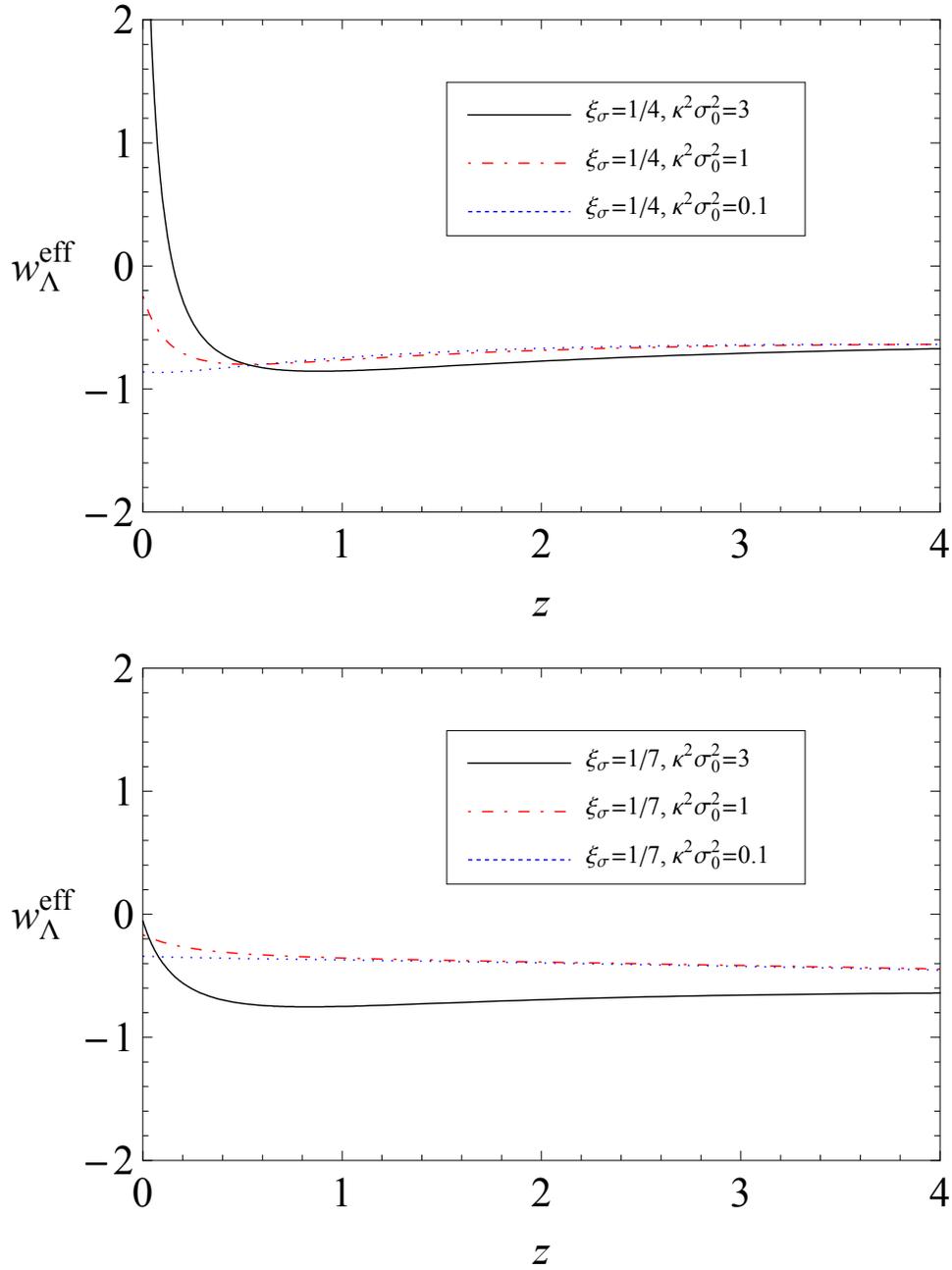}
\end{center}

\caption{{\it  $w_{\rm \Lambda}^{eff}$ vs $z$ in the interacting
phantom field case, for $\xi_{\sigma}=1/4$, $\xi_{\sigma}=1/7$,
$b^{2}=0.01$ and $k=+1,0$, where in each case the combination
$\k^2\sigma_0^2$ is taken equal to $3$, $1$, $0.1$ respectively.
The curves for $k=+1,0$ appear superimposed showing that the
effect of the non-flat universe is negligible. The divergence of
$w_{\rm \Lambda}^{eff}$ is a direct consequence of the singularity
of (\ref{wphantom}), and thus the corresponding combinations of
and $\k^2\sigma_0^2$ must be excluded.}} \label{phantomfig}
\end{figure}

\begin{figure}[quintom]
\begin{center}
\includegraphics[width=.8\textwidth]{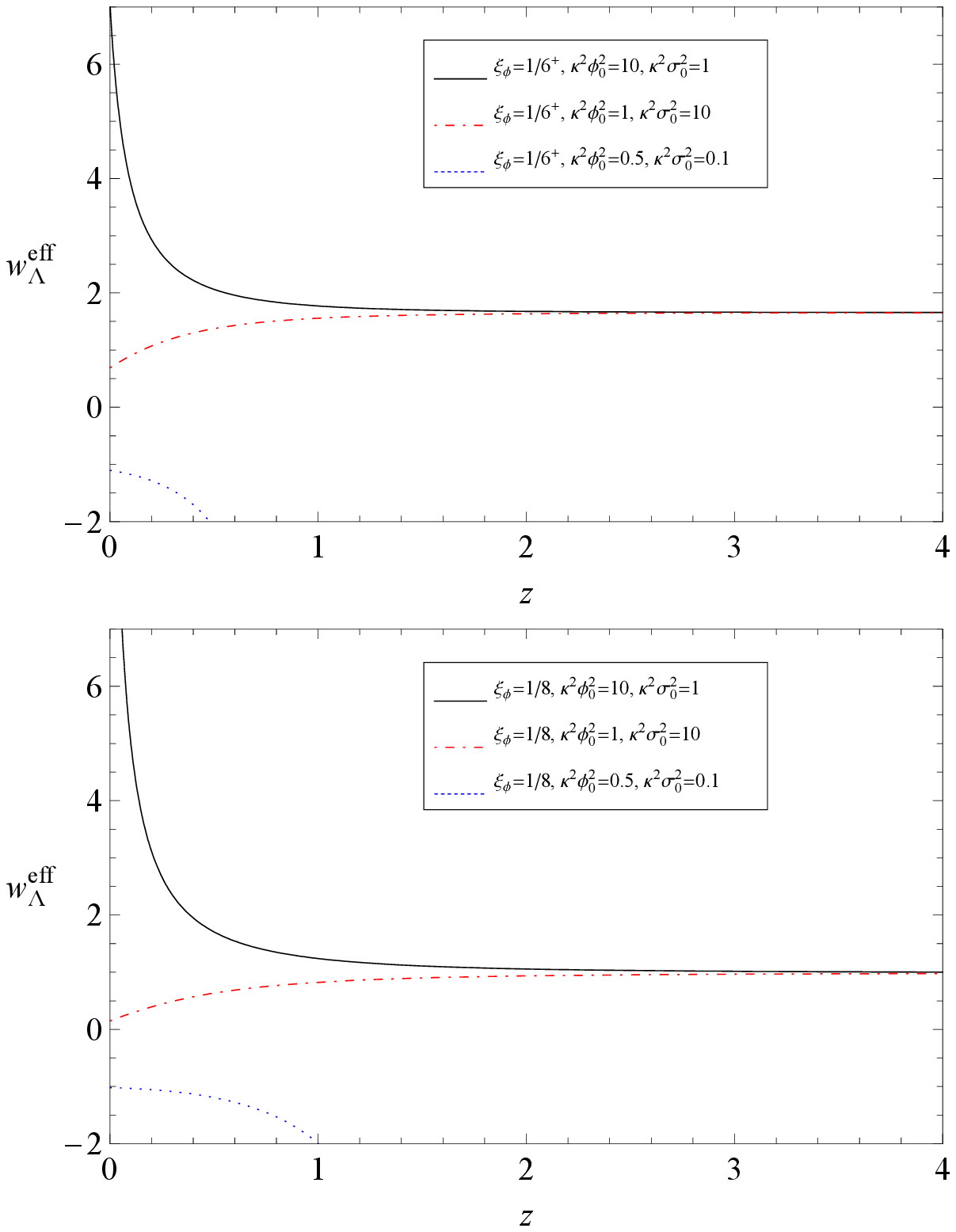}
\end{center}

\caption{{\it  $w_{\rm \Lambda}^{eff}$ vs $z$ in the combined
interacting quintom scenario, for $\xi_{\phi}=1/6^{+}$,
$\xi_{\phi}=1/8$, $b^{2}=0.01$ and $k=+1,0$, where in each case
the combinations $\k^2\phi_0^2$ and $\k^2\sigma_0^2$ are shown in
the insets. The curves for $k=+1,0$ appear superimposed showing
that the effect of the non-flat universe is negligible. The
divergence of $w_{\rm \Lambda}^{eff}$ is a direct consequence of
the singularity of (\ref{wquintom}), and thus the corresponding
combinations of $\xi_{\phi}$, $\k^2\phi_0^2$ and $\k^2\sigma_0^2$
must be excluded.}} \label{quintomfig}
\end{figure}
 \section{Conclusions}
Currently scalar fields play crucial roles in modern cosmology. In
the inflationary scenario they generate an exponential rate of
evolution of the universe as well as density fluctuations due to
the vacuum energy. It seems that the presence of a non-minimal
coupling (NMC) between scalar field and gravity is also necessary.
There are many theoretical evidences that suggest the
incorporation of an explicit NMC between the scalar field and
gravity in the action \cite{far}. The NMC arises from quantum
corrections and it is required also by the renormalization of the
corresponding field theory. Amazingly, it has been proven that the
phantom divide line crossing of dark energy described by a single
minimally coupled scalar field with general Lagrangian is even
unstable with respect to the cosmological perturbations realized
on the trajectories of the zero measure \cite{vik}. This fact has
motivated a lot of attempts to realize the crossing of the phantom
divide line by a equation of state parameter of the scalar field
used as dark
energy candidate in more complicated frameworks.\\
Studying the interaction between dark energy and ordinary matter
will open the possibility of detecting dark energy. It should be
pointed out that evidence was recently provided by the Abell
Cluster A586 in support of the interaction between dark energy and
dark matter \cite{bert}. However, despite the fact that numerous
works have been carried out, at present there are no stringent
observational bounds on the strength of this interaction
\cite{feng1}. This weakness to set stringent (observational or
theoretical) constraints on the strength of the coupling between
dark energy and dark matter stems from our unawareness of the
nature and origin of the dark components of the Universe. It is
therefore more than obvious that further work is needed in this
direction. Due to this, we have extended our work in \cite{set1}
to the interacting case in this paper. As a result, in the present
paper we have investigated canonical, phantom and quintom models,
with the various fields being non-minimally coupled to gravity, in
the framework of the interacting HDE in non-flat universe. For
this case, the characteristic length is no more the radius of the
event horizon ($R_E$) but the event horizon radius as measured
from the sphere of the horizon ($L$). In each case we have
extracted $w_{\rm \Lambda}^{eff}$, that is, the dark energy
effective equation of state parameter, as a function of the
redshift and used as parameters the couplings and the amplitudes
of the fields. Finally, we have analyzed it in order to obtain its
cosmological implications.
\section{Acknowledgment}The work of M. R. Setare has been supported by Research Institute for Astronomy
and Astrophysics of Maragha, Iran. The work of Alberto
Rozas-Fern\'{a}ndez was supported by DGICYT (Spain) under Research
Project No.~FIS2005-01180.

\end{document}